\documentclass[12pt,preprint]{aastex}

\newcommand{\be}{\begin{equation}}
\newcommand{\rhlh}{R_h/R_h(\hbox{heavy})}
\newcommand{\rhl}{\mbox{\ensuremath{R_h}}}
\newcommand{\rhh}{\mbox{\ensuremath{R_h(\hbox{heavy})}}}
\newcommand{\IMS}{{\em MS }}
\newcommand{\NOIMS}{{\em NOMS }}

\newcommand{\STD}{{\em S}}
\newcommand{\ltorder}{\hbox{ \rlap{\raise 0.425ex\hbox{$<$}}\lower
		      0.65ex\hbox{$\sim$} }}
\newcommand{\gtorder}{\hbox{ \rlap{\raise 0.425ex\hbox{$>$}}\lower
		      0.65ex\hbox{$\sim$} }}

\shorttitle{Mass Segregation}
\shortauthors{McMillan et al.}

\begin{document}

\title{A dynamical origin for early mass segregation in young star
  clusters}
\author{Stephen L. W. McMillan\altaffilmark{1}}
\author{Enrico Vesperini\altaffilmark{2}}
\affil{Department of Physics, Drexel University, Philadelphia, PA 19104}
\author{Simon F. Portegies Zwart\altaffilmark{3}}
\affil{Astronomical Institute `Anton Pannekoek' and Section
  Computational Science, \\ University of Amsterdam, Kruislaan 403, 
	1098SJ Amsterdam, the Netherlands}
\altaffiltext{1}{\tt steve@physics.drexel.edu}
\altaffiltext{2}{\tt vesperin@physics.drexel.edu}
\altaffiltext{3}{\tt spz@science.uva.nl}

\begin{abstract}
Some young star clusters show a degree of mass segregation that is
inconsistent with the effects of standard two-body relaxation from an
initially unsegregated system without substructure, in virial
equilibrium, and it is unclear whether current cluster formation
models can account for this degree of initial segregation in clusters
of significant mass.  In this Letter we demonstrate that mergers of
small clumps that are either initially mass segregated, or in which
mass segregation can be produced by two-body relaxation before they
merge, generically lead to larger systems which inherit the progenitor
clumps' segregation.  We conclude that clusters formed in this way are
naturally mass segregated, accounting for the anomalous observations
and suggesting that this process of prompt mass segregation due to
initial clumping should be taken fully into account in constructing
cluster dynamical models.
\end{abstract}

\keywords{star clusters: general}


\section{Introduction}

A population of massive stars of mass $m_h$ embedded in a cluster of
stars with mean mass $\langle m\rangle$ will sink toward the cluster
center on a time scale $t_{seg}\sim(\langle m\rangle/m_h) t_r$, where
$t_r$ is the cluster half-mass relaxation time scale (Spitzer 1987,
Binney \& Tremaine 1989).  This well known process of mass segregation
is a consequence of energy equipartition, whereby energy exchange
between stars by two-body relaxation causes more massive stars to slow
down and hence move inward in the cluster potential.  Strong
observational evidence for this mechanism has been found in many old
globular clusters (see e.g. Sosin 1997, Sosin \& King 1997, Koch et
al. 2004, Pasquali et al. 2004), consistent with the fact that these
systems have relaxation times significantly less than a Hubble time.

Interestingly, a number of studies also show significant mass
segregation in clusters having actual ages, as measured by the
evolutionary state of the component stars, substantially less than the
time needed to produce the observed segregation by standard two-body
relaxation (Hillenbrand 1997, Hillenbrand \& Hartmann 1998, Fischer et
al. 1998, de Grijs et al. 2002, Sirianni et al. 2002, Gouliermis et
al. 2004, Stolte et al 2006).  Numerical simulations indicate that
dynamical evolution from initially unsegregated systems cannot account
for the degree of mass segregation observed in these clusters
(e.g. Bonnell \& Davies 1998).

The obvious explanation is that these clusters were born mass
segregated, and recent theoretical studies do indeed suggest that
massive stars form preferentially in the centers of star-forming
regions (Klessen 2001, Bonnell et al. 2001, Bonnell \& Bate 2006).
The mechanism invoked to explain this primordial mass segregation
relies mainly on the higher accretion rate for stars in the centers of
young clusters.  However, the efficiency of this mechanism is still a
matter of debate (Krumholz et al. 2005, Klein \& McKee 2005, Bonnell
\& Bate 2006) and, more generally, the processes of massive star
formation and feedback remain poorly understood (Krumholz et
al. 2005).  Simulations of cluster formation have so far been confined
to small systems containing up to $\sim10^3 M_\odot$ in stars; it is
not currently known how the above findings scale to larger clusters.

In this Letter we report initial results of an extensive numerical
study exploring possible dynamical routes to mass segregation during
the early stages of cluster evolution.  Within the context just
described, we imagine that stars form in small clumps, which
subsequently merge to form larger systems (see e.g. Bonnell, Bate \&
Vine 2003 for models of hierarchical formation of star clusters; see
also Elmegreen 2006 and references therein).  We assume that the
clumps are either already significantly mass segregated at formation,
or small enough that mass segregation can occur within the merger time
scale.  In either case, we find that the final clusters inherit the
mass segregation of their progenitor clumps, providing a natural
mechanism for the production of larger systems which are mass
segregated yet physically young.  In \S2 we describe the approach and
initial conditions adopted in our investigation.  In \S3 we present
our results.  In \S4 we summarize our conclusions and briefly describe
further work now in progress.


\section{Method and Initial Conditions}

Our study is based on direct $N$-body simulations carried out using
the {\tt starlab} package\footnote{\tt http://www.manybody.org}
(Portegies Zwart et al. 2001), accelerated by the GRAPE-6
special-purpose hardware (Makino et al. 2003, Fukushige, Makino \&
Kawai 2005).

For all simulations presented here, we have adopted ``clumpy'' initial
conditions, in which the initial cluster consists of $N_c$ clumps,
their centers uniformly distributed within a sphere of radius
$R_{cluster}$.  (Note that we adopt terminology in which the cluster
initially consists of individual clumps; alternatively, we might
imagine a ``super star cluster'' initially consisting of individual
clusters.)  The system of clumps is not in virial equilibrium, and the
velocities of the clump centers are negligible (zero in the
simulations reported here).  Our simulations have $N_c = 2$ and 4, and
explore the evolution of systems having two specific values of the
``clumping ratio'' ${\cal R}_c \equiv R_{clump}/R_{cluster}$, where
$R_{clump}$ is taken to be the 90 percent Lagrangian radius of an
individual clump.  The two sets of runs have (i) ${\cal R}_c = 0.2$,
corresponding to clumps that are relatively close to one another and,
in some cases, effectively in contact, and (ii) ${\cal R}_c = 0.07$,
representing clumps that are widely separated.  Hereafter we refer to
these choices as ``moderately clumped'' and ``strongly clumped,''
respectively.

The individual clumps are modeled as systems of $10^4$ particles in
virial equilibrium, with Plummer density profiles (see e.g. Heggie \&
Hut 2003).  We present here the results of simulations with clumps
comprising just two components: $N_1$ light particles of mass $m_1$
and $N_2$ particles of mass $m_2$, with $m_2/m_1=20$ and
$N_1/N_2\simeq18$.  The numbers are chosen so that $m_2/\langle m
\rangle \approx10$.  While a two-component mass function is obviously
not representative of the IMF of a real cluster, this simplified IMF
allows us to focus our attention on the essential elements of mass
segregation dynamics for systems containing enough massive particles
to produce statistically significant results.  We note that the total
mass in the massive component is unrealistically high in this case,
but it yields significantly better statistics and does not affect our
overall conclusions; we return to this point below.

We describe two distinct sets of simulations.  In the first (hereafter
called the \IMS runs), we assume that the clumps are initially mass
segregated.  In the second (the \NOIMS runs) the individual clumps are
unsegregated---that is, both components are distributed with the same
half-mass radius.  Initial mass segregation in the \IMS runs is
achieved by first letting a representative (\NOIMS) clump evolve in
isolation for long enough for mass segregation to occur by normal
two-body relaxation.  We monitor the time evolution of the
``segregation factor'' $f_{seg} \equiv \rhlh$, where $\rhl$ and $\rhh$
are the half-mass radii of the entire cluster and of the heavy
component, respectively, and stop our calculation when $f_{seg}$
reaches an approximate steady state (see Figure 1).  This
mass-segregated system is then used as a template for all clumps in
our simulations.  We emphasize that this procedure is just a
convenient means of generating a self-consistent mass-segregated
system as an initial condition for an \IMS clump.

Table 1 summarizes the simulations described in \S3.  The \IMS
simulations are intended to explore the evolution of clusters produced
from individual clumps in which initial mass segregation has already
occurred due to processes acting at the time of star formation.  They
target the evolution of mass segregation during the merging process,
quantifying the mass segregation in the clusters resulting from such
clumpy systems.  The \NOIMS runs, on the other hand, explore whether
significant mass segregation can be produced in individual clumps
before they merge, and whether that mass segregation is preserved in
the final system.

\begin{table}[h]
\begin{center}
\begin{tabular}{|l|c|c|c|l|c|}
\hline
$~~~N$ & $N_c$ & mass seg. & clumping & $N_1/N_2$ & Figure\\
\hline
10,000 & 1 & \NOIMS & ---      & 18, 72 & 1 \\
20,000 & 2 & \IMS   & strong   & 18, 72 & 1 \\
40,000 & 2 & \IMS   & strong   & 18, 72 & 1 \\
\hline
40,000 & 4 & \IMS   & moderate & 18 & 2, 3  \\
    &   &        & strong   &    & 2, 3  \\
40,000 & 4 & \NOIMS & moderate & 18 & 3     \\
    &   &        & strong   &    & 3, 4  \\
40,000 & 4 & \NOIMS & strong   & (Kroupa MF) & 3     \\
\hline
40,000 & 1 & \NOIMS & ---      & 18 & 4 (model \STD) \\
\hline
\end{tabular}
\caption{Initial conditions of all simulations.  The columns list (1)
  the total number of particles in the run, (2) the number of clumps,
  (3) whether or not the clumps are mass segregated, (4) the degree of
  clumping, (5) the ratio of light to heavy particles (for the adopted
  mass ratio of $m_2/m_1=20$, a value of 18 here corresponds to a mass
  fraction of 53\% in the form of heavy particles; 72 corresponds to
  22\%), and (6) the relevant figure in the text.}\label{Table:ICs}
\end{center}
\end{table}


\section{Results}
\subsection{Systems with initial mass segregation (\IMS runs)}

The goal of the \IMS runs is to determine whether initial mass
segregation is preserved during the merger of two or more clumps and,
if so, to quantify it and establish a connection between the mass
segregation of the original clumps and that of the cluster resulting
from the merger.

Our first sets of simulations describe a ``hierarchical'' merger
scenario, in which two identical, initially mass-segregated clumps are
placed at a separation of ten times their half-mass radius with zero
relative velocity, and allowed to merge.  Subsequently, two copies of
the merged system are again placed at a separation of ten half-mass
radii with zero relative velocity and merged.  Figure 1 shows the time
evolution of $f_{seg}$ during these runs.  The two panels in the
figure correspond to different choices of the mass fraction of heavy
stars: the upper panel has $N_1/N_2=18$, corresponding to a heavy mass
fraction of 53\%, while $N_1/N_2=72$ in the lower panel, for a heavy
mass fraction of 22\%.  Both choices show qualitatively similar
results.

\begin{figure}
\epsscale{1}
\plotone{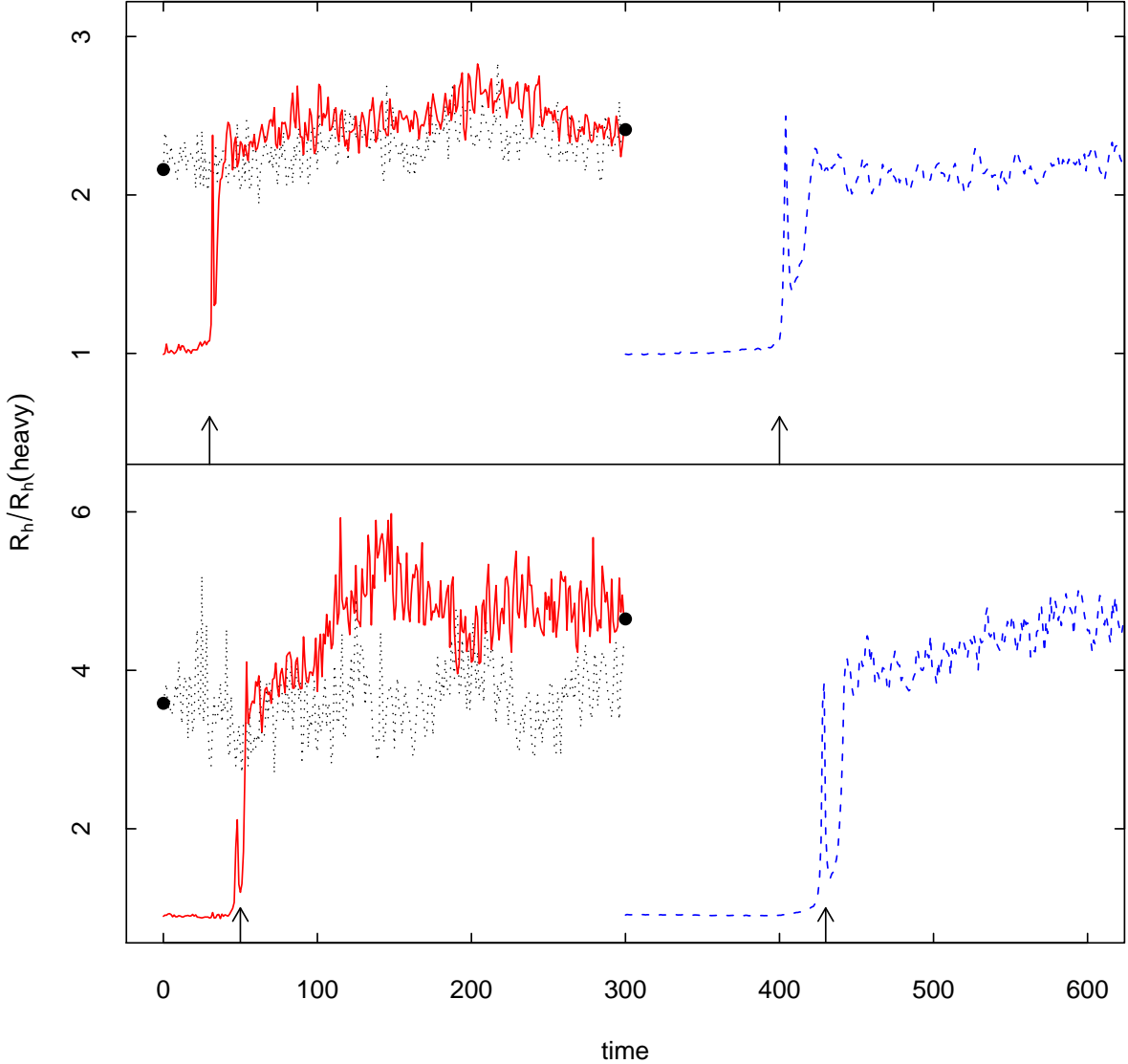}
\caption{Time evolution of the segregation factor $f_{seg} = \rhlh$,
measured relative to the system center of mass, for the hierarchical
merger simulations described in \S3.1.  The upper panel shows results
for $N_1/N_2=18$, the lower panel for a smaller fraction of heavy
particles ($N_1/N_2=72$).  Dotted lines show the variation of
$f_{seg}$ for the initial clumps used in the merging simulations, when
evolved in isolation.  Solid lines show the time evolution of
$f_{seg}$ when two such clumps merge.  Dashed lines refer to the
second merger simulation, in which two copies of the first merger
product are allowed to coalesce.  The dot at the end of each solid
line indicates the state of the clumps at the start of the second
merger calculation.  The vertical arrows mark the times when the
mergers occur.  Here, as in all figures, the time unit is the
dynamical time scale (Heggie \& Mathieu 1986) of one of the initial
unsegregated clumps used to generate the initial models.}
\label{fig1}
\end{figure}

In each panel, the first and second merger simulations are
represented by solid and dashed lines, respectively.  The approximate
times of the mergers themselves are indicated by arrows.  The dot at
the end of the solid line represents the starting time and initial
state of the second merger calculation.  Note that this figure
suggests no mass segregation at the start of each merger because the
individual clumps, while themselves segregated, are initially widely
separated, and the Lagrangian radii are measured relative to the
center of mass of the system.  The dotted lines show the evolution of
a single clump in isolation, indicating the value of $f_{seg}$ before
the merger and demonstrating that internal dynamics leads to
negligible structural evolution within each clump on the merger time
scale.  The dot at the left end of that line represents the initial
condition for individual mass-segregated clumps in the next set of
simulations.

In all cases, once the merger is complete, the amount of mass
segregation in the final cluster, as measured by $f_{seg}$, is
approximately equal to that in the original clumps---mass segregation
is preserved during the merging process.  This result is consistent
with the findings of van Albada (1982) and Funato, Makino \& Ebisuzaki
(1992), who showed that memory of particles' initial binding energy is
not erased during violent relaxation.

\begin{figure}
\epsscale{1}
\plotone{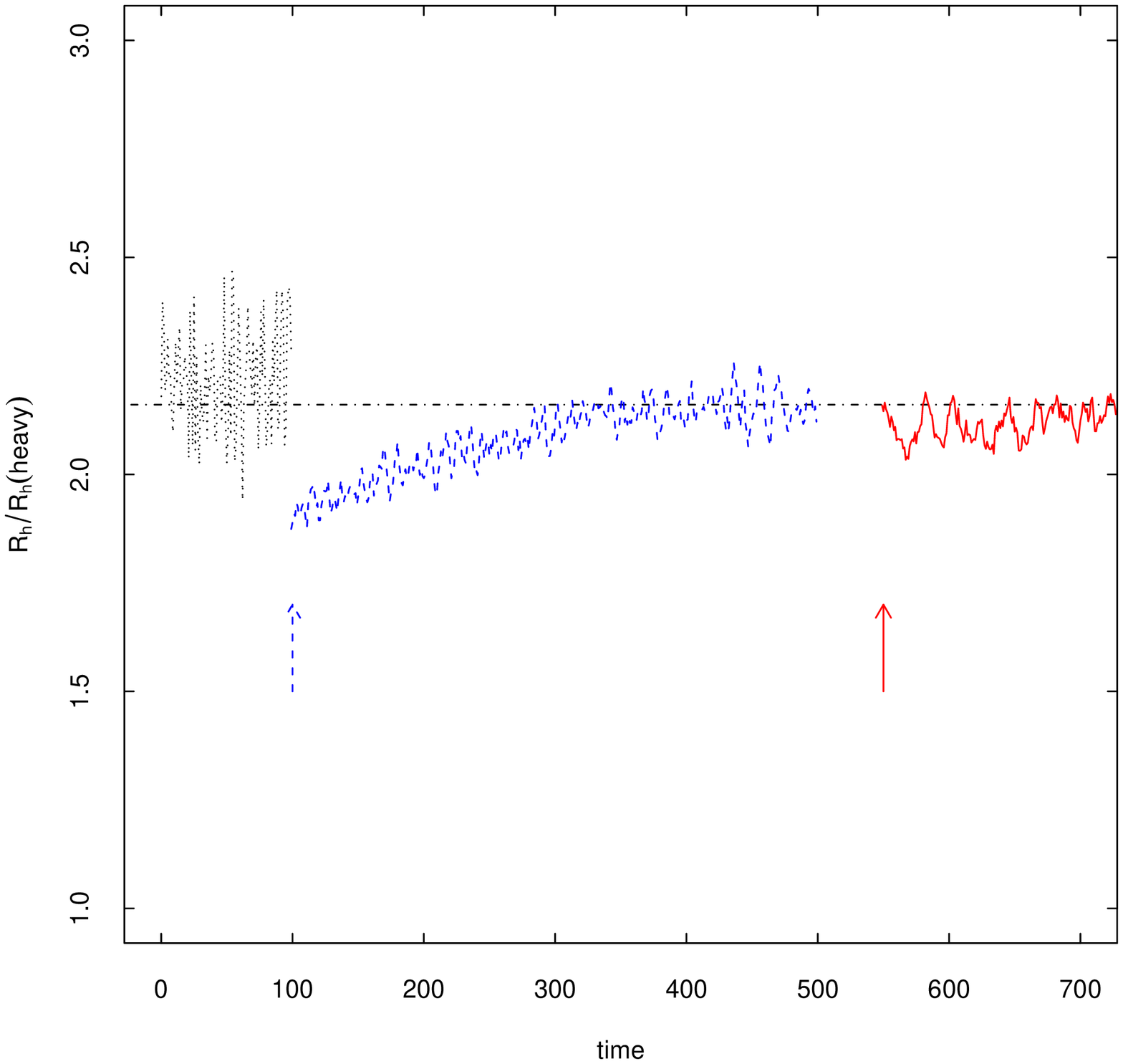}
\caption{Time evolution of $f_{seg}$ for simulations starting from 4
mass-segregated clumps (\IMS runs), for the two choices of the
clumping ratio described in the text.  The dashed line is for a
moderately clumped initial system; the solid line is for the strongly
clumped case.  The dashed (solid) vertical arrow marks the effective
end of the merging process for the moderately (strongly) clumped
simulation (Fig. 3 shows the duration of the merging process more
clearly).  The dotted line shows the time evolution of $f_{seg}$ for
the initial clumps used in the merging simulations, when evolved in
isolation; the horizontal dot-dashed line shows the initial value of
$f_{seg}$ in the individual clumps.}
\label{fig2}
\end{figure}

Figure 2 presents the time evolution of $f_{seg}$ for several \IMS
simulations with $N_c=4$.  The dashed line shows an average of three
random realizations of a moderately clumped system; the solid line
shows the average of two strongly clumped runs.  The dotted line at
left shows a portion of the internal evolution of an isolated
component clump (as in Figure 1, upper panel), as an indicator of the
state of the clumps before the merger occurs.  For clarity, we show
the behavior of $f_{seg}$ only after each merger is effectively
complete (vertical arrows), when the merged cluster is approaching
dynamical equilibrium and its center can be reliably defined.

We see that in these more general cases too, the cluster resulting
from the merger inherits the mass segregation of the component clumps:
violent relaxation during the merging process does not erase the
memory of the clumps' initial mass segregation.


\subsection{Initially unsegregated systems (\NOIMS runs)}

In order to explore whether initial mass segregation is an essential
ingredient in the scenario just described, we have repeated two of the
$N_c=4$ simulations described in the previous section, but without
initial mass segregation in the individual clumps.  Figure 3 compares
the time evolution of $f_{seg}$ for these simulations with the
corresponding \IMS simulations having the same initial distribution of
clumps.  In this case we follow the detailed merger history of the
original clumps, to illustrate how mass segregation proceeds first
within the clumps and then within each new merger product, culminating
in the final merged cluster.  In the moderately clumped case,
individual mergers start to occur quite rapidly, even before
significant internal mass segregation has occurred.  In the strongly
clumped case, we can clearly see internal mass segregation in some
clumps before they merge.  However, in both cases, the final values of
$f_{seg}$ are again comparable to those found in the \IMS simulations.

In the \NOIMS simulations, the segregation properties of the
end-products are largely controlled by the ratio of $t_{mrg}$, the
merging timescale of the cluster, to $t_{seg}$, the time required for
internal dynamics to produce mass-segregated systems used as initial
conditions in the \IMS runs.  For our choice of system parameters, we
find $t_{mrg}/t_{seg} \sim 0.3-0.5$ for moderately clumped initial
conditions, and $\sim 2-3$ for the strongly clumped case.  Thus, in
both cases, the merger occurs on a timescale longer than or comparable
to the mass segregation timescale of the individual clumps, so that
mass segregation occurs in the clumps before the clumps merge
completely, and the eventual degree of mass segregation is similar to
that found in the \IMS simulations.  As an additional point of
comparison, the dot-dashed line in Figure 3 illustrates that the
effect persists when a realistic (Kroupa et al. 1993) cluster mass
function is used.

\begin{figure}
\epsscale{0.9}
\plotone{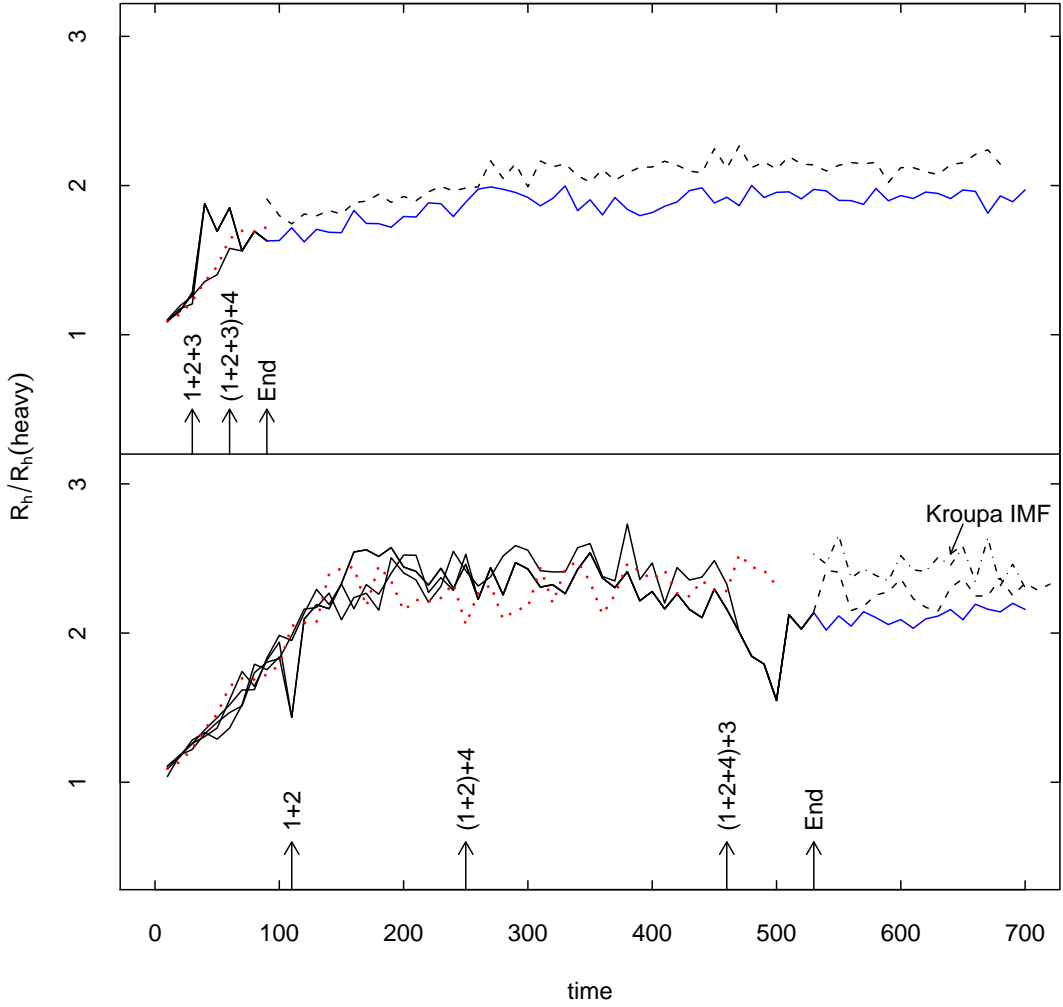}
\caption{Comparison of the time evolution of $f_{seg}$ for simulations
with initially mass-segregated clumps (\IMS runs, dashed lines), and
simulations starting with the same initial clump positions and
velocities but without initial mass segregation (\NOIMS runs, solid
lines).  The upper and lower panels show data for moderately and
strongly clumped initial conditions simulations, respectively.  For
clarity, only the final (post-merger) portions of the \IMS runs are
shown.  Vertical arrows mark various merging events between the clumps
(arbitrarily numbered 1--4 initially); the labels above each arrow
indicate the clumps involved in the merger.  The rightmost arrow in
each panel (labeled ``End'') marks the end of the merging process,
when the system is approaching dynamical equilibrium and has a single
well-defined center. The solid lines at each stage of the merging
process show the evolution of $f_{seg}$ for the remaining clumps in
the cluster.  The dotted lines show the evolution of $f_{seg}$ for an
individual clump evolved in isolation.  The dot-dashed line in the
lower panel shows the results of a comparable \NOIMS simulation with a
Kroupa initial mass function; in this case we plot the ratio of the
half-mass radius of the whole system to the half-mass radius of stars
having masses between 2.5 and 8 solar masses.}
\label{fig3}
\end{figure}

These simulations demonstrate an interesting and potentially important
alternative route leading to early mass segregation in young clusters.
Rather than relying on initial mass segregation in the clumps, this
scenario hinges on the multiscale nature of the cluster early
evolution: mass segregation is produced in individual small clumps,
and is approximately preserved by the subsequent merging.  We note
that, for the \IMS runs, the merging time is the only timescale
relevant to the process of forming a single large mass-segregated
cluster, as the segregated cores merge rapidly once the clumps come
into contact.  However, for the \NOIMS runs, the additional parameter
$t_{mrg}/t_{seg}$ plays a key role.  For this reason, the results of
the \NOIMS and \IMS runs are expected to scale differently with
increasing clump and/or cluster mass.


\section{Possible dynamical histories of young segregated clusters}

The end products of the simulations described above are young, yet
significantly mass segregated, clusters.  Without knowing the actual
dynamical history of the system, one might imagine ``observing'' one
of these simulated clusters to try to reproduce its properties and
reconstruct its past dynamical evolution.  The traditional way to do
this is to perform $N$-body simulations starting from the initial
conditions adopted in the vast majority of numerical studies of star
cluster evolution---a spherical system with no primordial mass
segregation and a Plummer (or King) density profile.  We have carried
out this experiment, running a simulation starting from a
two-component spherical system in virial equilibrium, with 40,000
particles and a Plummer density profile.  Hereafter we refer to this
simulation with standard initial conditions as model \STD.

Figure 4 (left frame) compares the time evolution of $f_{seg}$ in
model {\STD} with the strongly clumped \NOIMS run described in \S3.2.
As just discussed, mass segregation occurs much sooner (at least a
factor of $\sim 7-10$ faster) in the clumped case.  As shown in the
right frame, the density profiles at the ends of the two runs, when
the clusters exhibit approximately the same amount of mass
segregation, are very similar.  Since model {\STD} takes much longer
than the \NOIMS system age to reproduce the same cluster properties,
one might incorrectly conclude from this numerical study that the mass
segregation found in this cluster must reflect its initial conditions.
However, as we have shown, several possible dynamical histories can
lead to similar final systems.

\begin{figure}
\epsscale{1}
\plotone{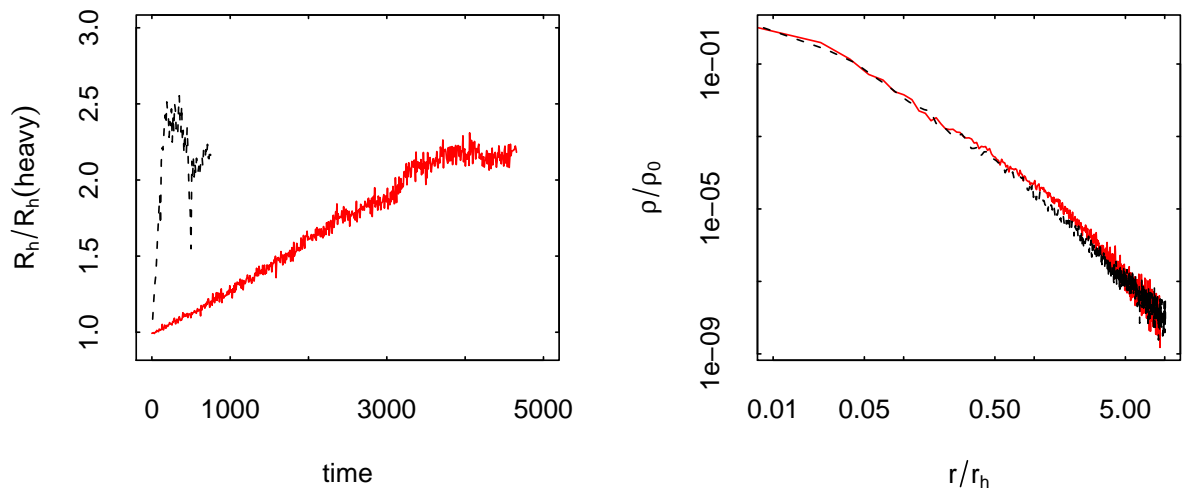}
\caption{(Left) Time evolution of $f_{seg}$ for the two simulations
discussed in \S4.  The left (dashed) curve began from strongly clumped
\NOIMS initial conditions with $N_c=4$ (see the lower panel of Figure
3); the right (solid) curve from a single unsegregated Plummer
profile.  (Right) The density profiles of the two runs at the end of
the simulations are almost indistinguishable.}
\label{fig4}
\end{figure}


\section{Conclusions}

We have presented the results of simulations following the early
evolution of star clusters, exploring the origin of mass segregation
observed in young clusters.  Our simulations started from clumpy
initial conditions with and without initial mass segregation in the
individual clumps, and studied the properties of the resulting merged
cluster.  Our main conclusions are:

\begin{enumerate}

\item For clumps with initial mass segregation, the degree of mass
  segregation in each clump is largely preserved during the merging
  process.

\item For clumps without initial mass segregation, the individual
  clumps may become mass segregated by two-body relaxation before the
  clumps merge.  In this case too, this mass segregation is
  subsequently inherited by the resultant merged cluster.

\item For clumped initial conditions, with or without initial mass
  segregation, the end-products of our simulations are young clusters
  whose properties are inconsistent with an initially unsegregated
  equilibrium cluster model.

\end{enumerate}
Our simulations demonstrate that there are a number of viable
evolutionary paths, relying on initial mass segregation in clumpy
systems and/or on multiscale dynamical evolution, that can lead to a
significant level of mass segregation in a physically young cluster.
(See also Vesperini et al. 2006 for a further dynamical mechanism
potentially leading to initial mass segregation.)

The results reported in this Letter cover only a small portion of the
possible initial conditions for the clumpy systems of interest here.
The full parameter space has many dimensions, spanning both the
properties of individual clumps (density profile, stellar IMF, number
of particles, virial ratio) and the properties of the larger system of
clumps (density profile, velocity distribution, clumping ratio, number
of clumps, clump mass distribution).  A systematic investigation of
the initial parameter space and the effects of early mass segregation
on the long term evolution of clusters is currently in progress; the
results will be presented elsewhere (Vesperini et al. 2006, in
preparation).


\section*{References}
\noindent
Binney, J., \& Tremaine, S. 1989, {\em Galactic Dynamics}, Princeton
  University Press\\
Bonnell, I.A., Clarke, C.J., Bate, M.R., \& Pringle, J.E.  2001,
  MNRAS, 324, 573\\
Bonnell, I.A., Bate, M.R., Vine S., 2003, MNRAS, 343, 413\\
Bonnell, I.A., Bate, M.R., 2006, astro-ph0604615\\
de Grijs, R., Gilmore, G.F., Johnson, R.A., \& Mackey, A.D. 2002,
  MNRAS, 331, 245\\
Elmegreen, B., 2006,  to
  appear in {\em Globular Clusters: Guides to Galaxies} (T.Richtler et al.,
  eds.) ESO/Springer (astro-ph0605519)\\ 
Fischer P., Pryor C., Murray S., Mateo M., \& Richtler T. 1998, AJ,
  331, 592\\
Fukushige, T., Makino, J.,  Kawai,A., 2005, PASJ, 57, 1009\\
Funato Y, Makino J., Ebisuzaki, T., 1992, PASJ, 44, 291\\
Gouliermis, D., Keller, S. C.; Kontizas, M.; Kontizas, E.;
  Bellas-Velidis, I., 2004 , A\&A, 416, 137\\
Heggie, D. C., \& Mathieu, R. D. 1986, in {\em The Use of
  Supercomputers in Stellar Dynamics} (P. Hut and S. McMillan, eds.;
  Springer-Verlag, New York)\\
Heggie, D., \& Hut, P. 2003, {\em The Gravitational Million-Body
  Problem}, Cambridge University Press\\
Hillenbrand L. A. , 1997, AJ, 113, 1733\\
Hillenbrand L. A., Hartmann L. E., 1998, ApJ, 331, 540.\\
Klessen R., 2001, ApJ, 556, 837\\ 
Koch, A., Grebel, E. K., Odenkirchen, M., Martínez-Delgado, D.,
  Caldwell,J. A. R. , 2004, AJ, 128, 2274\\
Kroupa,P., Tout,C.A., Gilmore,G., 1993, MNRAS, 262, 545\\
Krumholz M.R., Klein R.I., McKee C.F., 2005, Nature, 438, 332\\
Makino,J., Fukushige,T., Koga,M., Namura,K., 2003, PASJ, 55, 1163\\
Pasquali, A.,De Marchi, G., Pulone, L., Brigas, M. S., 2004, A\&A,
  428, 469\\
Portegies Zwart, S. F., McMillan, S. L. W., Hut, P., \& Makino,
  J. 2001, MNRAS, 321, 199\\
Sirianni, M., Nota, A., De Marchi, G., Leitherer, C., \& Clampin, M.
  2002, ApJ, 579, 275\\
Sosin, C., 1997, AJ, 114, 1517\\
Sosin, C., King, I.R., 1997, AJ, 113, 1328\\
Spitzer L., 1987, {\em Dynamical Evolution of Globular Clusters},
  Princeton Univ. Press\\
Stolte, A., Brandner, W., Brandl, B., Zinnecker H. 2006, AJ, 132, 253\\
van Albada, T.S., 1982, MNRAS, 201, 939\\
Vesperini, E., McMillan, S. L. W., \& Portegies Zwart, S. F. 2006, to
  appear in {\em Globular Clusters: Guides to Galaxies} (T.Richtler et al.,
  eds.) ESO/Springer\\

\end{document}